\documentclass[showpacs,preprint,amsmath,superscriptaddress,pre]{revtex4}  
\usepackage[dvips]{graphicx}
\usepackage{tabularx}
\usepackage{epsfig}
\usepackage{color}
\usepackage{subfigure}
\usepackage{soul}

\begin{document}

\title{
Finite-time scaling for epidemic processes
with power-law superspreading events
}
\author{Carles Falc\'o}
\affiliation{%
Centre de Recerca Matem\`atica,
Edifici C, Campus Bellaterra,
E-08193 Barcelona, Spain
}\affiliation{Departament de Matem\`atiques,
Facultat de Ci\`encies,
Universitat Aut\`onoma de Barcelona,
E-08193 Barcelona, Spain
}
\affiliation{Present address:
Mathematical Institute, University of Oxford, OX2 6GG Oxford, United Kingdom}
\author{\'Alvaro Corral}
\affiliation{%
Centre de Recerca Matem\`atica,
Edifici C, Campus Bellaterra,
E-08193 Barcelona, Spain
}\affiliation{Departament de Matem\`atiques,
Facultat de Ci\`encies,
Universitat Aut\`onoma de Barcelona,
E-08193 Barcelona, Spain
}
\affiliation{Complexity Science Hub Vienna,
Josefst\"adter Stra$\beta$e 39,
1080 Vienna,
Austria
}

\begin{abstract}
Epidemics unfold by means of a spreading process from
each infected individual to a variable number of secondary cases.
{It has been claimed
that the so-called superspreading events in COVID-19
are governed by a power-law-tailed distribution of secondary cases,
with no finite variance.}
Using a continuous-time branching process,
we demonstrate that for such power-law-tailed superspreading
the survival probability of an outbreak
as a function of both time
and
the basic reproductive number
fulfills a ``finite-time scaling'' law
(analogous to finite-size scaling)
with universal-like characteristics only dependent on
the power-law exponent.
{This clearly shows
how the phase transition separating
a subcritical and a supercritical phase
emerges
in the infinite-time limit (analogous to the thermodynamic limit).}
%
{We also
quantify the counterintuitive hazards
this superspreading poses.
When the expected number of infected individuals is
computed removing extinct outbreaks,
we find a constant value in the subcritical phase
and a superlinear power-law growth in the critical phase.
}

%






%
%
\end{abstract}

\date{\today}

\maketitle

\section{Introduction} 
The ongoing COVID-19 pandemic has
raised considerable concern over
the superspreading phenomenon.
In the propagation of infectious diseases, superspreading refers to when a single infected individual triggers a large number of secondary cases.
Superspreading has been previously proposed to happen in diseases
such as SARS \cite{SARS}, MERS \cite{MERS}, measles \cite{Lloyd_superspreading}, and the Ebola virus disease \cite{Ebola}.
Naturally, understanding
superspreading is crucial
not only for identifing which events drive the propagation but
also
for implementing effective contention measures \cite{Lewis_superspreading,Sneppen_superspreading_prl,modelling_covid19}.

Most definitions of superspreading have been rather vague or arbitrary.
For instance, some authors may define a superspreading event if a single individual provokes the direct contagion of at least 10 other individuals (secondary cases) \cite{Kucharski_book,Lewis_superspreading}.
Lloyd-Smith et al. \cite{Lloyd_superspreading}
associated the phenomenon to the presence of outliers in the distribution of secondary cases when these are modeled in terms of a Poisson distribution (with a mean given by the empirically-found value of the basic reproductive number $R_0$).
Thus, an excess of outliers would suggest that the Poisson distribution is not appropriate,
and a negative binomial is introduced instead
\cite{Lloyd_superspreading,endo_abbott_kucharski,Hebert_Dufresne}
(this
{comprises Poisson as a particular case
and
arises as a mixture of
Poisson secondary cases with gamma-distributed rates for different individuals).}

In other instances, superspreading has been associated to the 20/80 rule
\cite{Galvani_May},
in which the top 20\% of most infective individuals are the direct cause of a very large percentage of direct transmission (e.g., 80\%).
But note that the fulfilment of the 20/80 rule, providing a single pair of numbers, is not sufficient to characterize a probability distribution. In concrete, although for precise values of its parameters the negative binomial distribution fulfils the rule, other distributions may be tuned to fulfil the rule as well (for instance, the power law \cite{Newman_05}).
In summary,
the common approaches to superspreading
identify it with a distribution of secondary cases that has a large variance
\cite{Hill_epidemics}
(or at least larger than that of a Poisson distribution).


{Recent empirical observations of SARS-CoV and SARS-CoV-2 transmission show that
superspreading makes
the tail of the distribution of secondary cases
in these diseases
incompatible with an exponential tail
(which characterizes the negative binomial).
Instead, the decay is consistent with a power-law tail
\cite{Wong_superspreading}
{with an exponent $\gamma$ in between 2 and 3
for the probability mass function $p_k$}}
\cite{footnote_carles1,poisson_mixture,footnote_carles3}.
%
%
A fundamental difference between
both types of distributions
is that power-law-tailed distributions
(with such a value of $\gamma$)
cannot be characterized by their variance,
which diverges.
In this context, the mean, $R_0$,
is of limited applicability,
as a standard error cannot be associated to it
and variability becomes infinite,
{making the value of $R_0$ difficult to constrain empirically
and making extremal superspreading events probable occurrences}.
In an abuse of language,
we will refer to ``power-law superspreading''
when dealing with power-law tails with exponent in the range $2<\gamma<3$.
Although the empirical evidence supporting {power-law superspreading} could be weak,
we can speculate that
the power-law scenario makes sense in the light of both our knowledge of human social behavior
and the airborne transmission of COVID-19 \cite{Jimenez_lancet}
{(airborne transmission can skyrocket the number of secondary cases in poorly ventilated indoor spaces)}.



In this Article we explore the mathematical consequences of a power-law distribution of
secondary cases, as claimed for COVID-19 in Ref. \cite{Wong_superspreading}.
In this way,
we show that a
simple branching-process model
teaches important lessons to understand
spreading in infectious diseases
and its degree of universality,
in particular
regarding {power-law superspreading}.
The advantage of using simple model is that these can be very transparent
to show universal features, whereas more complicated models can be
of little practical use if their parameters cannot be precisely constrained
\cite{Castro_Ares_PNAS20}.

Although most used epidemic models are of the compartmental type
(or are based on compartments)
\cite{Hill_epidemics,Meyers,Pastor_rmp,Arenas_covid},
branching processes are well-known in the field
\cite{Lipsitch,Lloyd_superspreading,Kleinberg_book,Miller_pgf,pausch2021noise},
and are more convenient
to deal with stochasticity
(stochasticity is fundamental when there is large variability,
and superspreading is all about large variability),
and when it is required
to count the number of individual cases.

Also, branching process can approximate more complicated stochastic models \cite{Allen}
and are closely related to well-studied epidemic models on random networks \cite{KenahEpidemicsNetworks,BhamidiNam}.
Indeed, the equivalence between epidemic percolation networks and branching processes has been considered before \cite{kenahEpidemicPercolationNetworks},
and it is known that under certain conditions
both models predict the same
probability of epidemic, 
outbreak size distribution,
and
epidemic threshold
-- at least during the initial spread of the disease \cite{Kenah2007NetworkbasedAO}.
Recently, different types of branching-process models have been applied to study COVID-19
\cite{Riou,Bertozzi,Levesque,BP_immigration}.
Further, network models with power-law distributed number of connections
have been extensively studied, e.g. in Refs. \cite{Pastor_Vespignani,Castellano_Pastor,Pastor_rmp}.
%
%


First, we introduce the well-known continuous-time branching process;
then,
we study it for the infinite-variance case
using
a rather general family of power-law-tailed distributions.
We find that a finite-time scaling law provides
a universal description of power-law superspreading in terms of a unique scaling function
that is independent on model parameters.
{The finite-time scaling illustrates how
a continuous phase transition separating a subcritical and a supercritical phase
only emerges in the limit of infinite time
(which plays the role of a thermodynamic limit).}
Next, we compare with the case of finite-variance spreading,
with some counterintuitive results arising in the comparison.
Finally, we identify and quantify the hazard potentially arising from power-law superspreading.

%


\section{Preliminaries}  
We consider the age-dependent branching process with exponential lifetimes, also known as continuous-time branching process
\cite{branching_biology}.
At $t=0$ an initial element is created.
After an exponentially distributed lifetime,
the element generates a random number $k$ of offspring elements
and is removed from the population.
The new elements evolve in the same stochastic way,
each with an identical and independent exponential distribution of lifetimes,
with rate $\lambda$,
and an identical and independent distribution of offspring,
given by the probability $p_k$
(with $k=0, 1, \dots$).

{The branching-process assumption takes from granted
a well-mixed
and infinite susceptible population
(thus, one only needs to care about infected individuals),
as well as totally independent secondary cases.
{There are no sources of heterogeneity other than those coming from the offspring distribution.}
Note that the time dynamics given by the exponential lifetimes is different
from the way time progression is incorporated in network epidemic models
\cite{Noel,Allen_Allard}.
The higher-order structure of social interactions has been taken into account in recent
models \cite{StOnge}.
}

In the epidemic-spread analogy,
elements are infected individuals,
offspring are secondary cases,
and the removal of individuals at the end of their lifetime
corresponds either to recovery or death.
The total number of cases (secondary and beyond)
triggered by the initial infected individual will constitute an epidemic outbreak.
In usual approaches,
the offspring distribution $p_k$ can be given by a Poisson distribution, but,
as we have mentioned,
the negative binomial has been used to account for superspreading
\cite{Lloyd_superspreading}.
In contrast,
as it has been recently proposed,
we will consider
$p_k$ as a power-law-tailed distribution \cite{Wong_superspreading}.

The offspring distribution is characterized by its probability generating function (pgf), $f(s)=\sum_{k=0}^\infty p_k s^k$.
The mean (expected number of secondary cases, which is the basic reproductive number) is obtained
as $R_0=\langle k \rangle=f'(s)|_{s=1}$ (the prime denotes derivative).
The key (random) variable is $Z(t)$,
which counts the number of infected individuals at time $t$.
Its pgf $F(s,t)$ obeys
\begin{equation}
\frac{\partial F(s,t)}{\partial t} = \lambda\left[
f(F(s,t))-F(s,t)
\right],
\label{Fts}
\end{equation}
with initial condition $F(s,0)=s$ (at $t=0$ there is one single element)
\cite{branching_biology}.

Derivation of $F(s,t)$ with respect $s$ and taking $s=1$ yields $\mu(t)=\langle Z(t)\rangle$,
the expected number of elements (infected individuals) at $t$, fulfilling
$d\mu(t)/dt = \lambda (R_0-1) \mu(t)$,
with initial condition $\mu(0)=1$
{(we will refer to $\mu(t)$ as the mean instantaneous size of the outbreak, or just size).}
Straightforward integration leads to
$\mu(t) = e^{(R_0-1)\lambda t},$
which is decreasing if $R_0<1$
and increasing if $R_0>1$.
The case $R_0=1$ corresponds to the critical point
(see below).
It is remarkable
that the offspring distribution has null influence on $\mu(t)$,
except for its mean value $R_0$.
In other words, superspreading effects, no matter how they are defined
(from negative binomials or from power laws)
do not change the behavior,
as long as $R_0$
takes the required value.

The reason for this is that the mean number of infections
does not tell the whole story (only an averaged story).
To proceed, we need to calculate $\eta(t)$,
the probability that the outbreak is extinct at time $t$,
i.e., the probability of $Z(t)=0$.
As $\eta(t)=F(0,t)$, we only need to take $s=0$ in the equation for $F(s,t)$,
which leads to
\begin{equation}
\frac{d \eta(t)}{d t} = \lambda\left[
f(\eta(t))-\eta(t)
\right],
\label{ladeeta}
\end{equation}
with $\eta(0)=0$ \cite{branching_biology}.
As in the Galton-Watson (discrete-time) model \cite{Harris_original,branching_biology,Corral_FontClos},
the equation has a stable fixed-point solution, $\eta^*$, fulfilling $\eta^*=f(\eta^*)$,
and $f'(\eta^*)\le 1$.
Note that in the equation for $\eta(t)$,
the offspring pgf appears explicitly.

\section{Power-law-tailed offspring distributions}  
{Although {different definitions have been proposed \cite{Voitalov_krioukov},
one can simply} consider power-law-tailed (plt) distributions as those that behave asymptotically as a power law,
i.e.,
$p_k k^\gamma\xrightarrow[k\rightarrow \infty]{}c$
for
$c>0$ and an exponent $2<\gamma<3$ ensuring infinite variance.
{Power-law-tailed distributions defined in this way constitute
a subclass of the so-called regularly-varying distributions
(or, roughly speaking, fat-tailed distributions,
which, in extreme-value theory,
correspond to the so-called Fr\'echet maximum domain
of attraction \cite{Voitalov_krioukov}).}
See the Appendix for other possibilities.}

{In order to find the expansion of the pgf $f_\text{plt}(s)$ of $p_k$
we look at 
$f_\text{plt}''(s) = \sum_{k=0}^\infty k(k-1)p_k s^{k-2}$.
Note that $f_\text{plt}''(s)$ is well-defined for $0\leq s<1$ and diverges as $s\rightarrow 1$
{(divergence of the second moment),}
and also $k(k-1)p_k\sim k^{2-\gamma}$ for large $k$. By an Abelian theorem
\cite{aspects_random_walk}
{(applicable when $\gamma-2 < 1$)},
$f_\text{plt}''(s)$ behaves as $c\Gamma(3-\gamma)/(1-s)^{3-\gamma}$ near $s = 1$.
Integrating twice and using that $f_\text{plt}(1) = 1$ and $f_\text{plt}'(1) = R_0$,
we can write $f_\text{plt}(1-\epsilon)\approx
1-R_0 \epsilon +c \Gamma(1-\gamma) \epsilon^{\gamma-1}$ for small $\epsilon$.}


%

Now
we are able to find the probability of extinction from Eq. (\ref{ladeeta})
when this is close to one.
Let us introduce the survival probability of the outbreak at time $t$,
which is $q(t)=1-\eta(t)$.
Notice that the survival probability is the survivor function
of the outbreak lifetime
(i.e., a complementary cumulative distribution function,
but referring to outbreaks, not individuals,
and thus $\eta(t)$ is the corresponding cumulative distribution function).

For long times, and
close to the critical point
(which separates sure extinction
for $R_0\le 1$
from
a small probability of survival for $R_0>1$),
$\eta(t)$ will be close to one
and $q(t)$ will be close to zero.
So, we will be able to
apply in Eq. (\ref{ladeeta}) the previous expansion of the pgf
around $\eta(t)=1$ (i.e., $q(t)=0$) to get
$$
\frac{dq(t)}{dt}=\lambda\left[
(R_0-1) q(t) - {c \Gamma(1-\gamma)} q(t)^{\gamma-1}
\right],
$$
disregarding terms $\mathcal{O}(q(t)^2)$.
As we cannot apply the original initial condition,
because the equation is not valid for short times,
we substitute it for $q(t_0)=q_0$, with $q_0$ unknown.
The resulting solution
is
\cite{wolframalpha}
\begin{equation}
q(t)=
\left(
\frac{e^{(\gamma-2)(R_0-1)\lambda \Delta t}
(R_0-1)/[c \Gamma(1-\gamma)]}
{e^{(\gamma-2)(R_0-1)\lambda \Delta t}-1+
q_0^{2-\gamma}(R_0-1)/[c \Gamma(1-\gamma)]}\right)
^{\frac1{\gamma-2}}
\label{lasolucion}
\end{equation}
with $\Delta t=t-t_0$.

\section{Finite-time scaling}  
Close to the critical point, the solution verifies a finite-time scaling law
(analogous to finite-size scaling replacing system size by time
\cite{GarciaMillan,Corral_garciamillan}).
Defining the rescaled variable
\begin{equation}
z=(\gamma-2)(R_0-1)\lambda t,
\label{z}
\end{equation}
with $t-t_0 \simeq t$,
and disregarding the last term in the denominator of Eq. (\ref{lasolucion})
(which can be done close to the critical point, equivalent to long times when $z$ is finite)
we can write
\begin{equation}
q(t)=\left(\frac{1}{(\gamma-2) c \Gamma(1-\gamma)\lambda t}\right)^{\frac 1 {\gamma-2}} G_\gamma(z)
\,
\propto \frac{G_\gamma(z)}{t^{ 1 /{(\gamma-2)}}}
\label{scalinglaw2}
\end{equation}
with the $\gamma-$dependent scaling function
\begin{equation}
G_\gamma(z)=\left(\frac {z e^z}{e^z-1}\right)^{\frac 1 {\gamma-2}},
\label{scalingfunc}
\end{equation}
and where the dependence on the unknown initial condition has disappeared.
{Note that the new variable $z$, Eq. (\ref{z}),
absorbs in a rescaled way both the temporal dependence
and the distance to the critical point
(thus, in the forthcoming equations, time dependence is included both in $t$ and $z$).}

\begin{figure}[ht]
\includegraphics[width=0.5\columnwidth]{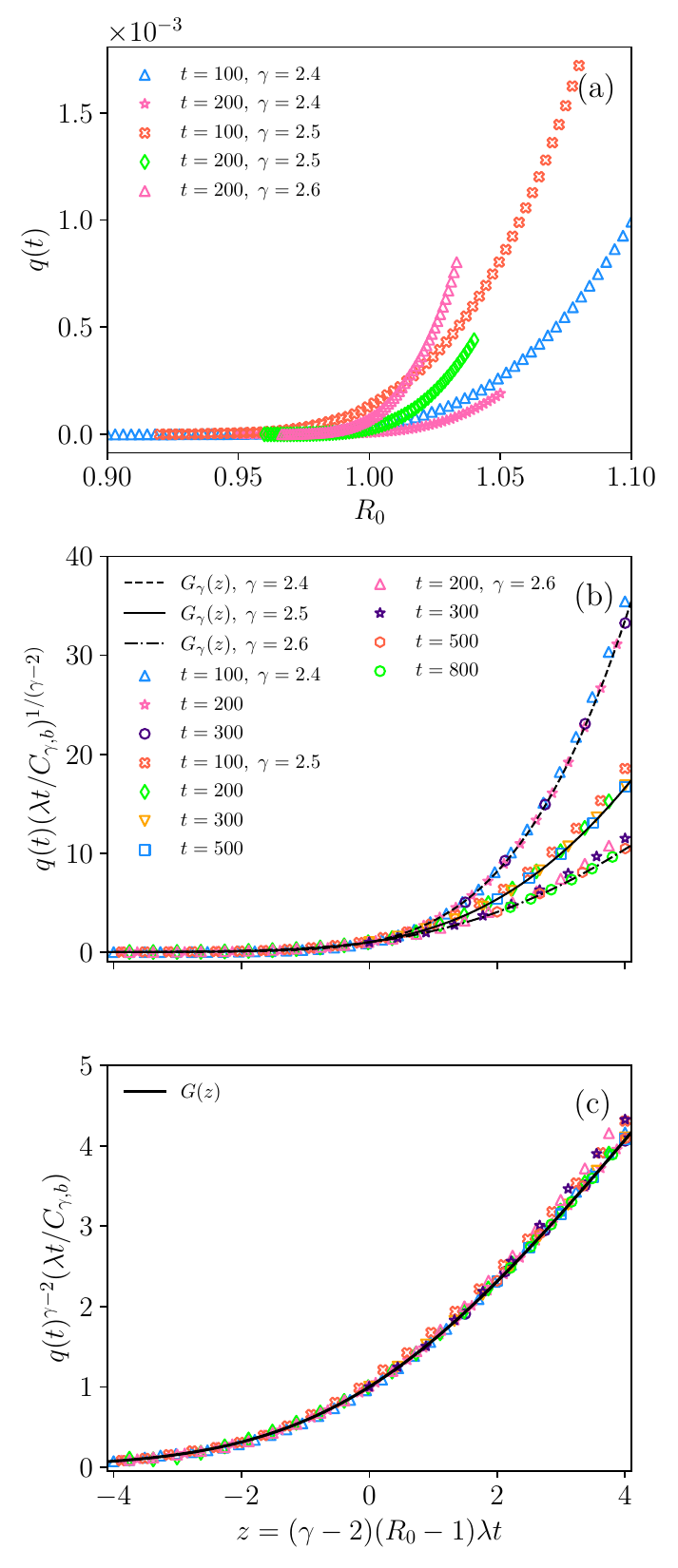}
\caption{
{
Results of computer simulations of the continuous-time branching process
with offspring distribution given by the shifted power-law distribution
(see Appendix),
for different values values of $b$ (to sweep $R_0$)
and different values of $\gamma$ and $t$, with $\lambda = 1$.}
{(a) Survival probabilities $q(t)$ versus $R_0$.
(b) Simple rescaling of $q(t)$ versus the rescaled ``distance'' to the critical point $z$.
(c) General rescaling of $q(t)^{\gamma-2}$.
Note that
$C_{\gamma,b} = \tfrac{\zeta(\gamma,b)}{(\gamma-2) \Gamma(1-\gamma)}$.
}
}
\label{fig_empirical}
\end{figure}

Therefore, for a fixed exponent $\gamma$,
displaying $(c \lambda t )^{1/(\gamma-2)} q(t)$
versus $z$ yields a unique $z-$dependent curve
independent of $\lambda$, $t$, $R_0$,
and any other parameter of the offspring distribution
(as long as $z$ is kept constant).
Further, for different values of $\gamma$,
displaying
$[(\gamma-2) c \Gamma(1-\gamma)\lambda t ] q(t)^{\gamma-2}$
versus $z$ the curve becomes additionally independent of
$\gamma$,
and therefore, ``universal,''
with $\gamma-$independent scaling function
{$G(z) = [G_\gamma(z)]^{\gamma-2}$.
}
The universal $\gamma-$independent scaling law is
\begin{equation}
q(t)^{\gamma-2}=\frac{1}{(\gamma-2) c\Gamma(1-\gamma) \lambda t}\, G(z)
\propto \frac 1 t \, G(z).
\label{scalinglaw3}
\end{equation}
The data collapses in Fig. \ref{fig_empirical},
obtained from computer simulations,
show how these finite-time scalings work.

The limiting behavior of the scaling function is
\begin{equation}
G_{\gamma}(z) \rightarrow
\left\{
\begin{array}{lcl}
\left(|z| e^{-|z|}\right)^{1/(\gamma-2)}
&\mbox{for} & z\rightarrow -\infty,\\
1 &\mbox{for} & z=0, \\
z^{1/(\gamma-2)} &\mbox{for} & z\rightarrow \infty\\
\end{array}
\right.
\label{limitingscalingfunc}
\end{equation}
{(it can be interesting to compare the resulting exponential decay for $q(t)$
in the subcritical phase with the empirical findings of Ref. \cite{Pastor_Vespignani}).}
Using this limiting behavior in the scaling law, the asymptotics of $q(t)$
(limit $t\rightarrow \infty$, close to the critical point)
becomes
\begin{equation}
q(t) \xrightarrow[t\rightarrow \infty]{}
\left\{
\begin{array}{lcl}
0 &\mbox{for} & R_0<1, \\
\left[(\gamma-2)c \Gamma(1-\gamma) \lambda t \right]
^{-1/(\gamma-2)}
&\mbox{for} & R_0=1, \\
\left[(R_0-1)/(c \Gamma(1-\gamma))\right]
^{1/(\gamma-2)}
&\mbox{for} & R_0>1. \\
\end{array}
\right.
\label{asymptotics}
\end{equation}

{This change of behavior at the critical value $R_0=1$ can be understood
as a phase transition (and $R_0=1$ as a critical point),
with $R_0$ the control parameter and
$\lim_{t\rightarrow\infty} q(t)$ the order parameter,
and with the asymptotic limit playing the role of the thermodynamic limit
{(infinite-system-size limit).}
This shows how the phase transition emerges when $t\rightarrow \infty$.
As $2< \gamma < 3$, the order-parameter exponent
$\beta=1/(\gamma-2)$
is in the range
$1 < \beta < \infty$
and the transition is not sharp
but continuous with a continuous first derivative at $R_0=1$.
So, the order of the transition is higher than second
(in contrast to the finite-variance case, see below).
The result for the critical phase is in agreement with
Ref. \cite{Saichev_Sornette_branching} for a discrete-time branching process.
An equivalent result to the one for the supercritical phase is known
in the context of percolation in scale-free networks \cite{Cohen_Havlin}.}
{


The close-to-critical regime, for which our results are valid,
can be of great practical interest, as
spontaneous changes in human behavior and
implementation of contention measures
usually
lead to a decrease in the value of $R$ \cite{Corral_epidemics_pre,Manrubia_Zanette}
(the equivalent of $R_0$ when this changes its value).}

\section{Comparison with the finite-variance case}  
The result for the case of finite variance
\cite{Wei_Pruessner_comment}
(Poisson, negative binomial, etc., but also power-law tail with $\gamma > 3$)
can be considered included in the previous expressions.
Indeed,
taking
Eqs. (\ref{z}), (\ref{scalingfunc}), (\ref{scalinglaw2}) and (\ref{asymptotics})
and replacing
$\gamma-2$ by 1
and
$c\Gamma(1-\gamma)$ by $\sigma^2/2$,
with $\sigma^2$ the variance of the offspring distribution in the critical point,
one recovers the formulas for the finite-variance case \cite{Wei_Pruessner_comment}.

{Thus,
the power-law behavior
of the order parameter as a function of $R_0$
in the case of infinite variance, Eq. (\ref{asymptotics}),
translates into a linear function 
in the case of finite variance,}
i.e.,
$q(t) \xrightarrow[t\rightarrow \infty]{} {2(R_0-1)}/{\sigma^2}$,
for $R_0>1$.
This 
highlights the importance of determining not only
the mean $R_0$ of the offspring distribution, but also its variance
(when it is finite \cite{Hebert_Dufresne}).
The problem with using the Poisson distribution for offspring is that the variance is
equal to $R_0$ and, close to the critical point, both are close to one.
{But there is nothing special with regard the negative binomial,
apart of allowing a variance different than $R_0$;
any distribution with the same variance and $R_0$ would lead
not only to the same asymptotic solution for $q(t)$
but to the same finite-time scaling law \cite{Wei_Pruessner_comment}.
In other words, superspreading with finite variance does not lead
to any new phenomenology.
It is only for power-law superspreading (with infinite variance)
that superspreading becomes a new phenomenon,
in the sense that new universality classes arise.
}

The previous simple expression for the limit of $q(t)$ in the finite-variance case
(together with $q(t) \rightarrow 0$ for $R_0\le 1$)
corresponds to the usual transcritical bifurcation
\cite{Corral_Alseda_Sardanyes}.
Nevertheless,
the power-law case with $2<\gamma<3$ also corresponds to a transcritical bifurcation,
despite the fact the behavior in the supercritical phase is not linear.
Comparing,
{for the same values of $R_0$},
the supercritical phases for finite and infinite variances,
one can see that,
sufficiently close to the critical point,
the linear term is above the nonlinear one,
and therefore the probability $q(t)$ that an outbreak does not get extinct
is smaller if there is {power-law superspreading}
(in fact, this probability is zero at first order in $R_0-1$,
in comparison with the finite-variance case).
Thus, power-law superspreading makes extinction of the outbreaks easier.


{We can quantify the differences in the outbreak lifetimes $t$.}
This is a random variable
with survivor function $q(t)$.
Although we have only calculated the tail of $q(t)$, 
this is enough to characterize the expected lifetime $\langle t \rangle$ of an outbreak.
From Eqs. (\ref{limitingscalingfunc})
and (\ref{asymptotics}),
it is clear that {in the infinite-variance case}
$\langle t \rangle$ is finite for $R_0< 1$
(because $q(t)$ decays exponentially)
and infinite for $R_0>1$
(because $q(t)$ does not tend to zero, and therefore it has a non-zero mass at infinity).
{This is valid also for finite-variance offspring distributions.}

The qualitative behavior at the critical point $R_0=1$ is different and counter-intuitive. 
In both cases we have critical slowing down (power-law decay in time),
but
in the finite-variance case
$q(t) \sim 1/t$, which means that
{the power-law exponent of the density is 2 and}
$\langle t \rangle$ diverges,
whereas for infinite variance, $q(t) \sim 1/t^{1/(\gamma-2)}$,
leading to an exponent
{of the density} larger than
two
and therefore to a finite mean value $\langle t \rangle$.
In other words, in the critical phase,
spreading with finite variance leads
(despite the probability of extinction is one)
to never-ending outbreaks (in expected value, not in single realizations),
but {infinite-variance superspreading} reduces the expected lifetime to be finite.

In any case,
power-law-tailed outbreak lifetimes
(or total outbreak sizes \cite{Kucharski_book,Cirillo_Taleb,Corral_epidemics_pre,Corral_comment_CT2})
are not an indication of {power-law superspreading},
as in the critical point power laws arise with any sort of spreading,
{whereas outside the critical point
power-law lifetimes do not take place, whatever the spreading.
It is important then not to confuse these two different types of power laws.
And of course, the occurrence of large outbreaks is not an indication of superspreading
{(they may arise even for the Poisson distribution if $R_0\ge 0$)}.}

%


\section{Hazard from power-law superspreading}  
We have seen that the expected number of infected individuals
varies exponentially as
$\mu(t)=e^{(R_0-1)\lambda t}$,
independently of the spreading characteristics,
but the survival probability of an outbreak is decreased when there is {power-law superspreading}.
Which are the hazards coming from this, then?
Obviously, $\mu(t)$ is not highly informative as it contains the contribution from
outbreaks that have got extinct (and contribute with a value of zero, but are counted).

In a formula, $\mu(t)=q(t) I_\text{sur}(t)+\eta(t) \times 0$,
with $I_\text{sur}(t)$ the expected number of infected individuals for outbreaks that are not extinct
at time $t$
(note that $I_\text{sur}(t)$ is an average between infected individuals, in contrast to $\mu(t)$);
therefore $I_\text{sur}(t)=\mu(t)/q(t)$,
and the decrease in $q(t)$ for {power-law superspreading} will yield an increase in $I_\text{sur}(t)$,
in concrete, substituting the scaling law for $q(t)$ [Eq. (\ref{scalinglaw2})]
{we get another finite-time scaling law},
\begin{equation}
I_\text{sur}(t)=
\left[{c\Gamma(1-\gamma)}{(\gamma-2)\lambda t}\,\left(\frac{e^z-1}z\right)\right]^{\frac 1 {\gamma-2}}
\xrightarrow[t\rightarrow \infty]{}
\left\{
\begin{array}{lcl}
\left(\frac{c\Gamma(1-\gamma)}{1-R_0}\right)^{1/(\gamma-2)}
&\mbox{for} & R_0<1, \\
\left[(\gamma-2)c\Gamma(1-\gamma)\lambda t\right]^{1/(\gamma-2)}
&\mbox{for} & R_0=1, \\
\left(\frac{c\Gamma(1-\gamma)}{R_0-1}\right)^{1/(\gamma-2)} e^{(R_0-1)\lambda t}
&\mbox{for} & R_0>1. \\
\end{array}
\right.
\label{peculiar}
\end{equation}
(the case of finite variance is recovered with the substitutions
$c\Gamma(1-\gamma) \rightarrow \sigma^2/2$ and
$\gamma-2 \rightarrow 1$).

These results mean that,
in the subcritical phase,
the very few outbreaks that survive reach a fixed average (instantaneous) size
$I_\text{sur}$
(while they survive \cite{footnote_carles4}),
{with a higher $I_\text{sur}$ in the case of power-law superspreading
(in comparison with the finite-variance case).}
In contrast, in the supercritical phase the non-extinct outbreaks grow exponentially
(with a prefactor that can be very high when $\gamma$ is close to 2).
{It is at the critical point that one finds
an important qualitative difference between
the finite-variance case and the power-law case:
in the former case the average size of the outbreaks that survive
diverges linearly, but for power-law superspreading the growth is superlinear
(as a power law of $t$ with exponent larger than one).
}
%
We note then a trivial yet important observational bias,
due to the fact that, at time $t$, we only see the outbreaks that have not become extinct.
This is a dramatic realization of the survivorship bias
(where survivorship refers to the outbreak, not to the individuals).

\section{Discussion}  

{We have made clear how a continuous-time branching process with power-law-tailed
secondary cases (in correspondence with recent observational results describing superspreading in COVID-19 \cite{Wong_superspreading})
has properties that are qualitatively different
to the case of finite variance.
The latter constitute a well-known mean-field universality class
with order-parameter exponent $\beta=1$ \cite{GarciaMillan_Universality},
whereas the power-law superspreading leads to a continuous of universality classes
depending on the value of the secondary-case power-law exponent $\gamma$ \cite{footnote_carles5}.}
Further, we derive the existence of a finite-time scaling law
describing the probability of outbreak survival
as a function of $R_0$ and time,
Eqs. (\ref{scalinglaw2}) and (\ref{scalinglaw3}),
and calculate the exact value of the scaling functions,
Eq. (\ref{scalingfunc}).
These scaling laws could be extended to random networks.
{We also show the peculiar behavior of $I_\text{sur}(t)$, Eq. (\ref{peculiar}).}


It would be desirable to apply our results to the COVID-19 pandemic,
in order to obtain the probability of outbreak extinction after some time
as a function of $R_0$.
for which the offspring power-law exponent $\gamma$ has been estimated.
However, in addition to the exponent $\gamma$,
knowledge of the constant $c$ in the asymptotic power-law formula is also fundamental
(a relation between $c$ and $R_0$ exists, but it is model dependent).
In other words, it is not enough to know the distribution of secondary cases for large $k$
\cite{Wong_superspreading},
but one needs to know the whole ``population'' to which those large outbreaks belong.
Thus, concentrating only in large outbreaks is useless for the calculation
of the survival probability.





\begin{acknowledgments}
We appreciate feedback from
M. Bogu\~n\'a,
L. Lacasa and R. Pastor-Satorras.
%
%
Support from projects
PGC-FIS2018-099629-B-I00
from the Spanish MICINN,
CEX2020-001084-M from
the Spanish State Research Agency,
through the Severo Ochoa and Mar\'{\i}a de Maeztu Program for Centers and Units of Excellence in R\&D,
as well as the CERCA Programme/Generalitat de Catalunya
is acknowledged.
\end{acknowledgments}

\appendix

\section{Shifted power-law offspring distribution}
One particular case of a power-law tailed distribution is given by the shifted power-law distribution. This is given by
\begin{equation}
p_k=\frac{1}{\zeta(\gamma,b) (b+k)^\gamma},
\end{equation}
for $k=0$, $1$,$\dots$
with $\gamma$ the exponent,
$b$ a location parameter,
and $\zeta(\gamma,b)$ the Hurwitz zeta function, ensuring normalization.
$\gamma >2$ leads to a finite mean
and $\gamma<3$ leads to an infinite variance.
This is the range we consider, together with $b>0$.
The mean of the distribution can be calculated directly to be
$R_0=\zeta(\gamma-1,b)/\zeta(\gamma,b)-b$.
The case $b=0$, truncated from below at $k=1$,
would lead to the standard discrete power law
(straight in a log-log plot),
whereas $b>0$ leads to a shifted discrete power law.
The shift does not alter the power-law behavior at the tail,
i.e., $p_k \sim 1/k^\gamma$, for any $b>0$ and large $k$.

Using the notation in the main text, in this case we have $c = 1/\zeta(\gamma,b)$. However, this simple case admits an alternative but equivalent analysis. The pgf of the shifted power law (pl) is
$f_\text{pl}(s)={\Phi(s,\gamma,b)}/{\zeta(\gamma,b)}$,
with $\Phi(s,\gamma,b)=\sum_{k=0}^\infty s^k/(b+k)^\gamma$
the so-called Lerch transcendent
and $\zeta(\gamma,b)=\Phi(1,\gamma,b)$.
When $b = 1$ one has
$\zeta(\gamma,1)=\zeta(\gamma)$ (the Riemann zeta function)
and
${\Phi(s,\gamma,1)}=\text{Li}_\gamma(s)/s$,
with $\text{Li}_\gamma(s)$
the polylogarithm,
arising in integrals that appear in the study of the Bose-Einstein condensation and whose asymptotic behavior near $s = 1$
is well known.
A generalization of this is possible
\cite{lerch_transcendental_book},
yielding
$$
\Phi(s,\gamma,b)=\frac{1}{s^b}\left[\Gamma(1-\gamma)(-\ln s)^{\gamma-1} + \zeta(\gamma,b) + \zeta(\gamma-1,b) \ln s +\mathcal{O}(\ln^2 s)\right],
$$
valid for $|\ln s| < 2\pi$, with $b>0$ and $\gamma$ a positive non-integer;
$\Gamma()$ is the gamma function.
Since we are interested in $s$ close to but smaller than 1,
we can write $s=1-\epsilon$,
and then $s^{-b}\simeq 1+b\epsilon$
and $\ln s \simeq -\epsilon$; thus
$$
f_\text{pl}(1-\epsilon)=\frac{\Phi(1-\epsilon,\gamma,b)}{\zeta(\gamma,b)}=
1-R_0\epsilon + \frac{\Gamma(1-\gamma)}{\zeta(\gamma,b)} \epsilon^{\gamma-1}+\mathcal{O}(\epsilon^2),
$$
where we have assumed the range of interest, $2<\gamma<3$.
The rest of the calculation is identical to that in the main text,
just replacing $c$ by $1/\zeta(\gamma,b)$.

\section{Alternative power-law tail}
Our results also hold for distributions $p_k$ that have similar behavior to power-law tailed distributions but do not satisfy the condition
$p_k k^\gamma\xrightarrow[k\rightarrow \infty]{}c$.
For instance, one could consider distributions satisfying the alternative condition
\begin{equation}
\sum_{k=0}^\infty k^2 \left(p_k-\frac{c}{(k+1)^\gamma}\right) < \infty,
\label{condition2}
\end{equation}
for a given real positive constant $c$.
Note that this condition is in fact different
from the one used in the main text,
since there are distributions
satisfying this condition which do not tend in the limit to a power law;
e.g., any distribution obeying
$p_k\sim \big(1 + (-1)^k/2\big)k^{-\gamma}$ for large $k$
(the sum can be shown to converge using Leibniz's criterion for alternating series).
Conversely, there are also distributions
for which the limit of $p_k k^\gamma$ for $k\rightarrow\infty$ exists, but the condition above is not satisfied
(e.g., when $p_k\sim k^{-\gamma}(1+1/\log k)$ for large $k$, with $\gamma<3$).

{In order to find the expansion of the pgf $f_\text{plt}(s)$ of $p_k$
when $p_k$ verifies the condition given by Eq. (\ref{condition2}),
we define $g(s)=f_\text{plt}(s)-c \zeta(\gamma) f_\text{pl1} (s)$
with $\zeta(\gamma)$ the Riemann zeta function
and $f_\text{pl1}(s)$ the pgf of the shifted power law with $b=1$.
The condition above guarantees the existence of $g'(1)$ and $g''(1)$,
and then, $g(1)=1-c \zeta(\gamma)$ and
$g'(1)=R_0-c \zeta(\gamma)R_\text{pl1}$,
with $R_0$ the mean of $p_k$
and $R_\text{pl1}$ the mean of the shifted power law.
In this way we can write
$f_\text{plt}(1-\epsilon)=g(1-\epsilon) + c\zeta(\gamma)f_\text{pl1}(1-\epsilon)=
1-R_0 \epsilon +c \Gamma(1-\gamma) \epsilon^{\gamma-1}+\mathcal{O}(\epsilon^2)$, which
is identical to the pgf derived in the main text
(although the range of validity is different).


\begin{thebibliography}{10}
\bibitem{footnote_carles1}
For other datasets better fits have been found by using {Poisson} mixtures as
  distributions of secondary cases \cite{poisson_mixture}.
\bibitem{footnote_carles3}
Reference \cite{Wong_superspreading} reports an exponent between 1 and 2
  because it refers to the exponent $\gamma-1$ of the complementary cumulative
  distribution function.
\bibitem{footnote_carles4}
The existence of a constant, non-zero value of {$I_\text{sur}$} could explain
  the observed persistence of computer viruses with low values of {$R_0$}, see
  {Chap.} 6 of {Ref.} \cite{Kucharski_book}. {In} particular, this could be
  another way to look at the results of {Ref.} \cite{Pastor_Vespignani}.
\bibitem{footnote_carles5}
The situation is similar to the generalized central-limit theorem
  \cite{Bouchaud_Georges}, for which there is a unique universality class (the
  {Gaussian} distribution) when the initial variance is finite, but infinite
  classes (the {L\'evy} stable distributions) for diverging variance. {Also},
  thinning processes show a similar separation but between finite and infinite
  means \cite{Corral_csf} {(this sharp separation between finite and infinite
  first or second moments does not happen for extreme-value limit distributions
  \cite{Coles}).}
\end{thebibliography}
\end{document}